\documentclass{aa}  

\usepackage{graphicx}
\usepackage{txfonts,hyperref}
\newcommand{\mic}{\,{$\mu$m}}

\begin{document}

   \title{CASSISjuice: open-source pipeline and offline complete atlas of Spitzer/IRS staring observations}


   \author{V. Lebouteiller
          \inst{1}
          }

   \institute{Université Paris-Saclay, Université Paris Cité, CEA, CNRS, AIM, 91191, Gif-sur-Yvette, France\\
              \email{vianney.lebouteiller@cnrs.fr}
             }

   \date{Submitted to arXiv September 15, 2023}

 
  \abstract
   {Mid-infrared spectroscopy provides many important diagnostics on gas and dust features in a wide variety of astrophysical objects. In the JWST era, it is important to maintain a durable database of observations with previous facilities, for preparing new observations or completing existing ones. The Spitzer Infrared Spectrograph observed more than 20\,000 targets with wavelengths as low as $\sim5.2$\mic\ and as long as $\sim38.0$\mic, thereby complementing JWST/MIRI data for long wavelength diagnostics and providing overall invaluable diagnostics together with JWST or in view of future IR facilities. } 
   {In order to maximize the science output of Spitzer/IRS, the CASSIS atlas has provided reduced IRS spectra since 2011, extracting and selecting the best spectrum from various methods. We now present CASSISjuice, an offline version of the pipeline and atlas, adding several hundred sources that had never cleared the pipeline in order to make it complete for the first time. }
   {We updated the low- and high-resolution pipelines in order to be able to process every IRS staring mode observation (i.e., all observations but maps), and we also upgraded the high-resolution pipeline to version 2. The new pipeline also associates the pointings within ``cluster'' observations resulting in a single spectrum (possibly low- and high-resolution) per position and therefore overall a single CASSISjuice ID per targeted position. Distinct observations at the same position are considered to be independent. }
   {The initial repositories are hosted at Zenodo, providing the open-source pipeline code and the atlas itself with specific attention to producing the smallest dataset possible. Version controlled repositories are also available at GitLab, including Python notebooks to illustrate the offline manipulation of the full atlas. }
   {The offline CASSISjuice atlas is meant to facilitate the analysis of large samples and the identification of potentially interesting or relevant spectra. We encourage redistribution and reuse of the atlas following the ``Attribution-NonCommercial-ShareAlike'' license. Citation guidelines are unchanged with the two seminal papers presenting the low- and high-resolution pipelines \citep{Lebouteiller2011a,Lebouteiller2015a}, possibly the present arXiv-only paper as well and, if necessary, the code itself associated with the specific version released \citep{Lebouteiller2023c,Lebouteiller2023b}. }


   \maketitle
%

\section{Introduction}

Infrared spectroscopic observations performed with the Infrared Spectrograph (IRS; \citealt{Houck2004a}) onboard the Spitzer Space Telescope \citep{Werner2004a} have led to significant progress and numerous publications in various astrophysical fields. The IRS performed about $15\,400$ observations (``AORkey'') in staring mode (i.e., targeting single sources in two consequent ``nod'' positions of the detector, as opposed to maps).

\begin{figure*}
   \centering
   \includegraphics[width=18cm]{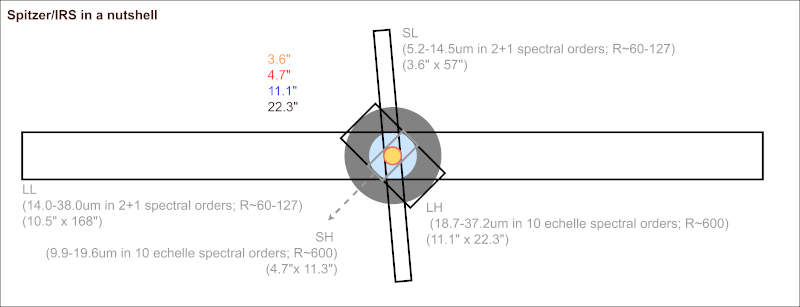}
      \caption{Low-resolution long slits and high-resolution apertures of the IRS. }
         \label{fig:nutshell}
       \end{figure*}

A fraction of these observations were done in ``cluster'' mode in which several pointings were targeted within a single observation, leading to a total of about $20\,400$ object positions and corresponding spectra. The observation setup includes all or a combination of the following:
\begin{itemize}
\item Low-resolution (LR) long-slits: Short-Low (SL) covering $5.2-14.5$\mic\ and Long-Low (LL) covering $14.0-38.0$\mic\ with a spectral resolving power $R\sim60-127$.
\item High-resolution (HR) apertures: Short-High (SH) covering $9.6-19.8$\mic\ and Long-High (LH) covering $18.7-37.2$\mic\ with $R\sim600$.
\end{itemize}
  Figure\,\ref{fig:nutshell} shows the aperture/slit size and relative orientations. The IRS data thus covers somewhat longer wavelengths than JWST/MIRI between $\approx28-37$\mic\ and includes sources that may be too bright or too extended for JWST. 

The immense legacy value of the IRS is well highlighted by the many studies still using the instrument data, even since 2009 and the end of the cryogenic mission phase. Part of this success is due to the availability of publishable-quality spectra for most observations through the CASSIS atlas, released in two parts with the low-resolution atlas \citep{Lebouteiller2011a} and the high-resolution one \citep{Lebouteiller2015a}. CASSIS provides automatically-selected products from various background-subtraction methods, spectral extraction methods, and flux calibration methods, along with many diagnostics to guide the user through the selection process. The spectra available through the web form (\href{http://cassis.sirtf.com}{http://cassis.sirtf.com}) and described in the above papers correspond to version 7 for low-resolution and version 1 for high-resolution.

The availability of publishable-quality IRS spectra enabled projects such as IDEOS (the Infrared Database of Extragalactic Observables with Spitzer; \citealt{HernanCaballero2016a,Spoon2022a}; \url{http://ideos.astro.cornell.edu/}). The IDEOS sample uses CASSIS to identify all extragalactic spectra obtained with the IRS low-resolution mode, with a particular attention to identifying misclassified objects and to selecting the best spectra when several observations of the same object are available. Other steps include the stitching of the spectra from different modules and the calculation of the redshift. 

Building upon CASSIS, CASSISjuice is meant to provide open-source, offline, access to the full IRS spectral atlas, with the goal in mind to facilitate the analysis of large samples and the identification of interesting/relevant spectra. A specific attention has been given to provide streamlined products in small repositories. 

CASSISjuice provides the LR and HR pipeline codes (IDL core programs + Python wrappers) through a public repository and also upgrades the HR pipeline to version 2 with several minor improvements. The full CASSISjuice atlas is provided for download through another public repository. The CASSISjuice atlas contains all IRS spectra performed in staring mode (i.e., not including maps), adding several hundred spectra compared to the CASSIS website, mostly due to pipeline fixes related to observations that did not go exactly as planned. Another important update concerns the organization of the pointings for cluster observations, leading to a single ID for each targeted position with corresponding LR and/or HR spectra (compared to previous versions of the atlas where IDs for LR and for HR were mixed). 

\noindent In summary, we provide:
\begin{itemize}
\item a public repository for the LR and HR pipeline codes which can be used for in-depth analysis of some spectral extractions and intermediary products, available as a version-controled repository (\url{https://gitlab.com/cassisjuice/pipeline}) including notebooks and as a citable repository corresponding to the present version (\url{https://zenodo.org/deposit/8339954}; \citealt{Lebouteiller2023b}),
\item a public repository with the CASSISjuice atlas that can be examined offline, containing the most important products as well as the CASSISjuice ``concentrate'' that includes the absolute minimal data set (i.e., the best spectra that were automatically selected by the pipeline), available as a version-controled repository (\url{https://gitlab.com/cassisjuice/atlas}) including notebooks and as a citable repository corresponding to the present version (\url{https://zenodo.org/record/8339965}; \citealt{Lebouteiller2023c}).
\end{itemize}

These repositories and the associated codes are released under the license ``Attribution-NonCommercial-ShareAlike 4.0 International'' (CC BY-NC-SA 4.0), i.e., it is possible to share (copy and redistribute the material in any medium or format) and adapt (remix, transform, and build upon the material) for non-commercial uses as long as proper credit is given and as long as the same license conditions propagate. Works using CASSISjuice data should cite the same reference publications: \cite{Lebouteiller2011a} and \cite{Lebouteiller2015a}. For reproductibility, we advise either simply mentioning the version number of the pipelines (currently v7 for LR and v2 for HR) or else adding the specific code citations \cite{Lebouteiller2023b} and \cite{Lebouteiller2023c} for the present versions. For significant or extensive use of the data, we also suggest citing the seminal Spitzer \citep{Werner2004a} and IRS papers \citep{Houck2004a}.

In the following, we briefly describe the main steps of the pipeline leading to the selection of the best methods, the atlas itself, and illustrations of possible applications.

\section{Pipeline summary}\label{sec:pipeline}

CASSISjuice performs many steps starting from the detector images downloaded from the Spitzer Heritage Archive at IRSA (\url{https://irsa.ipac.caltech.edu/applications/Spitzer/SHA/}): cleaning of bad pixels, combination of individual exposures, removal of the background (telescope + large-scale astrophysical emission), spectral extraction, combination of nod spectra, and flux calibration. Some additional steps include the removal of some artefacts either at the 2D or 1D level (e.g., fringes due to interferences from the light path within the instrument). We refer to \cite{Lebouteiller2011a,Lebouteiller2015a} for the details and describe in the following the most important steps. The pipeline extensively uses scripts from SMART \citep{Higdon2004a}, SMART/AdOpt \citep{Lebouteiller2010a}, IRSCLEAN (Ingalls 2011), and IRSFRINGE \citep{Lahuis2007a}.

\subsection{Background subtraction}\label{sec:bgsub}

\begin{figure}
   \centering
   \includegraphics[width=9cm]{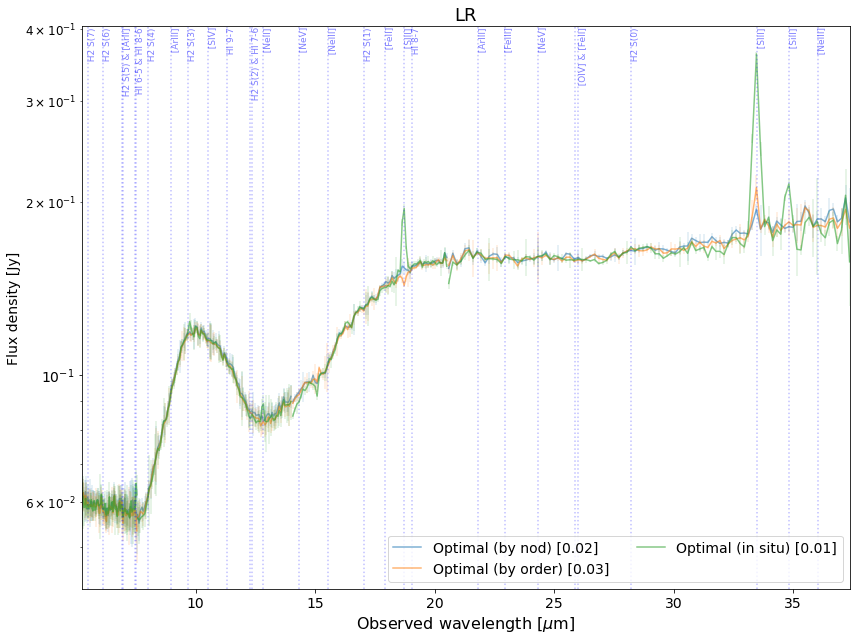}
   \includegraphics[width=9cm]{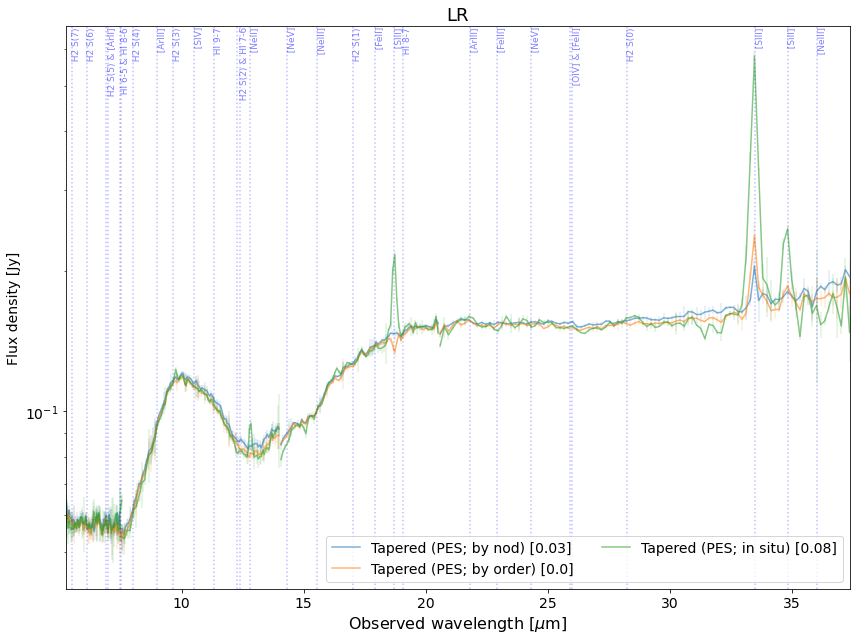}
      \caption{Comparing the background subtraction methods for optimal extraction (top) and tapered column extraction (bottom) for low-resolution mode (CASSISjuice ID 18147584\_0). Usually the subtraction "by nod" provides the best signal-to-noise ratio (SNR) compared to "by order" because the nod background lies in the same detector exposure. The "in situ" background subtraction usually leads to a worse SNR but it may be the only choice if contaminating sources are found in the offset detector images by nod or by order.}
         \label{fig:bgsub}
       \end{figure}

For low-resolution observations. the background can be removed either by subtracting the detector images corresponding to the other nod position (``by nod''), to another spectral order (``by order''), or by estimating the local continuum at the source location (``in situ''). The methods by nod/order ensure that the subtracted background corresponds exactly
to the same location in the detector, thereby not only removing the background emission but also mitigating rogue pixels. The methods by nod/order can be applied as long there is no contaminating source at the offset position. The various methods are compared to each other by the pipeline in terms of signal-to-noise ratio (SNR) and potential contamination by other sources in the offset positions (see a comparison in Fig.\,\ref{fig:bgsub}). 

The echelle-spectroscopy used for high-resolution observations make it impossible to subtract a clean image unless a dedicated offset background was observed. CASSISjuice does not associate observations with potential dedicated offset background positions and relies instead on other methods to remove the background. The first method consists in removing the local continuum simultaneously with the optimal extraction of the source (similar to the ``in situ'' method for low-resolution observations). While this effectively removes the large-scale physical emission, rogue pixels may still cause issues. The second method, applicable strictly to point sources, uses the differential spatial profile between the two nods and effectively mitigates rogue pixels (Sect.\,\ref{sec:extraction}).

\subsection{Spectral extraction}\label{sec:extraction}

\begin{figure}
   \centering
   \includegraphics[width=9cm]{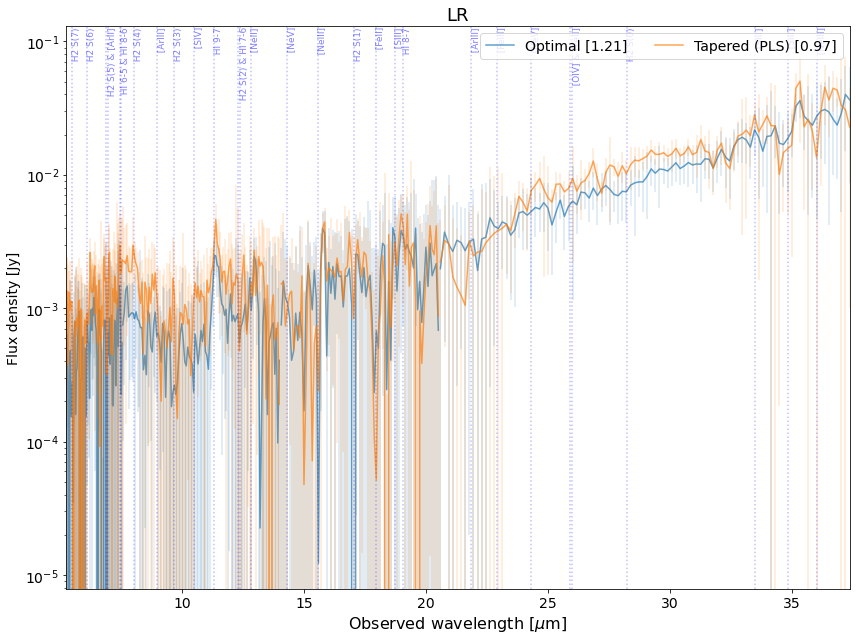}
   \includegraphics[width=9cm]{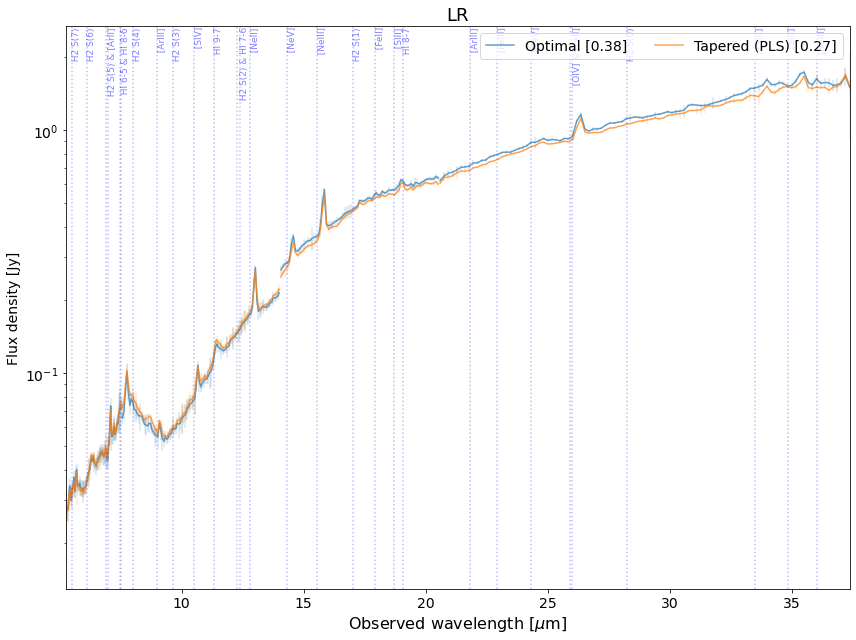}
      \caption{Comparing tapered column extraction (point-source calibration) to optimal extraction for a point source. Here we use the default background subtraction. Optimal extraction usualy leads to a larger signal-to-noise ratio (SNR) for faint sources (top; CASSISjuice ID 26864384\_2) while little difference is seen for bright sources (bottom; CASSISjuice ID 25408000\_0). }
         \label{fig:snr_lr}
       \end{figure}

\begin{figure}
   \centering
   \includegraphics[width=9cm]{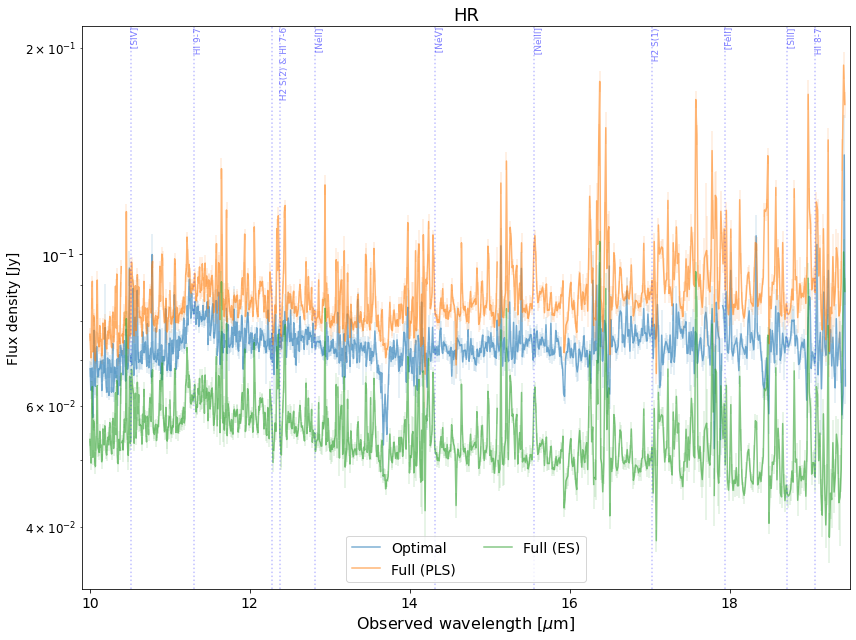}
      \caption{Comparing the optimal extraction to the full aperture extraction for the high-resolution mode, for CASSISjuice ID 25646848\_0. Full aperture extraction with point-source flux calibration can be used for point source but the signal-to-noise ratio (SNR) is usually best with optimal extraction. }
         \label{fig:snr_hr}
       \end{figure}

CASSISjuice then extracts the intended source around the requested position. The source may be point-like in which case the optimal extraction (using the point-spread function (PSF) profile as weights) provides the best SNR, or spatially-extended in which case a simple flux integration is performed along the cross-dispersion direction of the slit/aperture. Figures\,\ref{fig:snr_lr} and\ \ref{fig:snr_hr} illustrate potential differences between integrating the flux or using optimal extraction. In both cases (point source or extended source), optimal extraction is used anyway as a source finder algorithm to locate the source around the requested position and to identify potential sources in the offset images for background subtraction (Sect.\,\ref{sec:bgsub}). 

\begin{figure}
   \centering
   \includegraphics[width=9cm]{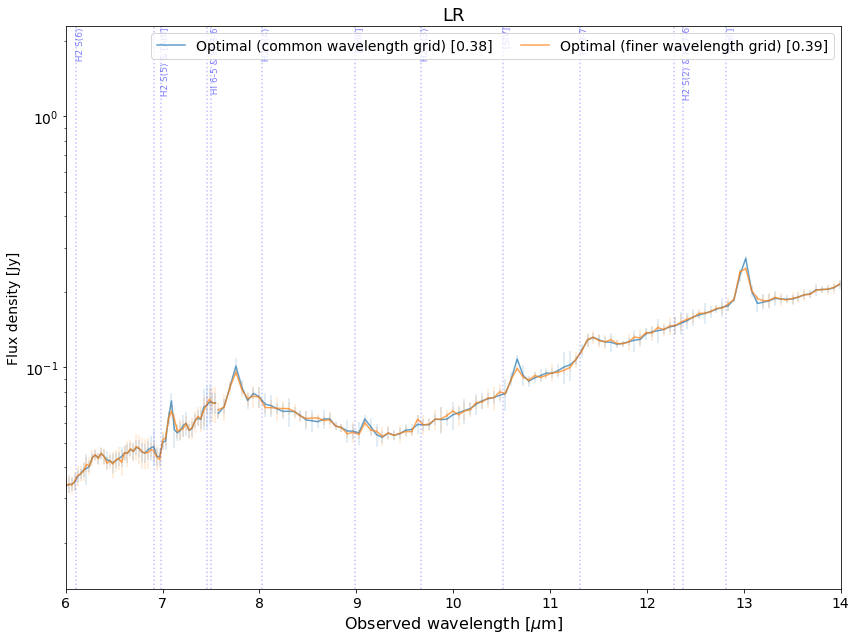}
   \caption{Comparing optimal extraction of both nods on common wavelength grid to both nods on their own wavelength grid (i.e., gaining a bit of spectral resolution) for CASSISjuice ID 25408000\_0. Using a common wavelength grid (slightly) optimizes the signal-to-noise ratio while using the finer wavelength grid (slightly) optimizes the spectral sampling and resolution.} 
         \label{fig:grid}
       \end{figure}

\begin{figure}
   \centering
   \includegraphics[width=9cm]{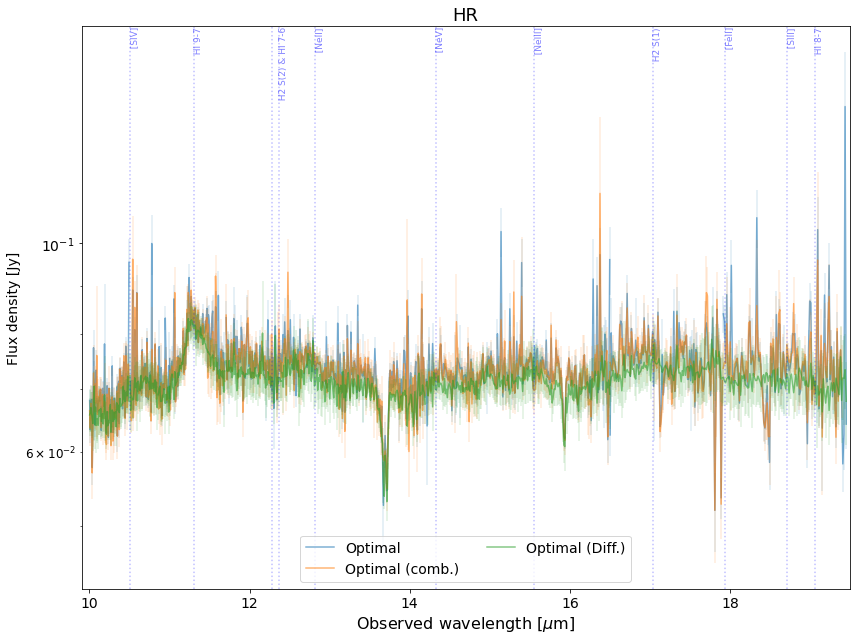}
      \caption{Comparing the various optimal extraction methods for CASSISjuice ID 125646848\_0: "nods" is the simultaneous extraction of both nods, "nodcomb" is the combination of the two nod spectra extracted separately, and "diff" is the extraction of the differential profile of nod 1 minus nod 2 (for pure point sources). The "nods" version is always the the default choice but the "diff" version can be checked to optimize the signal-to-noise ratio (although it is valid only for strict point sources).} 
         \label{fig:hr_opt}
       \end{figure}

If the source is point-like, several versions of the optimal extraction may be used depending on the mode. For low-resolution mode, the pipeline extracts the spectra at the two nod positions on their respective wavelength grid, thereby gaining a slight increase in spectral sampling and resolution if both spectra are interleaved, at at slight expense of SNR. Both versions (common wavelength grid or finer wavelength grid) are available (Fig.\,\ref{fig:grid}). For high-resolution mode, the pipeline can either extract 1) the two nods simultaneously leading to a single spectrum, 2) the two nods individually leading to two spectra that are then combined, or 3) the differential profile of the difference between the two nod detector images. In principle, the latter method provides the best SNR but may lead to systematic uncertainties if the source is not strictly point like (Fig.\,\ref{fig:hr_opt}). The flux calibration for optimal extraction is performed using theoretical and observed spectra of reference stars and does not require any kind of aperture corrections. 

Whether the source is point source or spatially extended, a ``tapered column'' extraction for low-resolution mode or full aperture for high-resolution mode can be used to integrate the flux. For point sources, the flux calibration accounts for the light lost outside the slit/aperture (especially in the short, dispersion direction) while for ``infinitely'' extended sources this correction is cancelled due to the fact that the fraction of light coming in/out the slit/aperture balances. If the source is neither a point source or an ``infinitely'' extended source (i.e., it is referred to as ``partially extended''), some attempts can be done to estimate a wavelength-dependent flux calibration that depends on the source spatial extent. Such a product is now proposed in CASSISjuice and is explicitly considered in the selection of the best spectrum. 

While other objects may be showing in the detector images apart from the requested object, corresponding to locations along the long slits or to locations within observations in another spectral order (low-resolution observations), there is currently no method to identify such sources apart from investigating manually the detector images at specific locations.

\subsection{Applications}

CASSISjuice provides the code for both LR and HR pipelines (written in IDL) through public repositories at GitLab and Zenodo as well as Python wrappers that were used to produce the atlas described in Section\,\ref{sec:atlas}. The general user is not expected to run the pipeline and should use the atlas -- described in the following -- instead. Expert users may find the pipeline availability useful to understand the spectral products and their automatic selection, to diagnose some issues, to access intermediary products, or even to improve some steps in the pipeline itself. More generally, the availability of the pipeline code indirectly ensures the durability of the IRS spectral atlas and consequently the long-term legacy of the data. In that vein, we release the code under a license permitting copies and adaptations as long as proper credit is given.

\section{Atlas}\label{sec:atlas}

Ultimately, the various methods presented in Section\,\ref{sec:pipeline} are compared and a selection is made to choose the best spectrum for the final atlas. The atlas contains the full dataset of IRS spectra observed in staring mode and is provided through other, specific repositories.

\subsection{Catalog}

\begin{figure*}
   \centering
   \includegraphics[width=18cm]{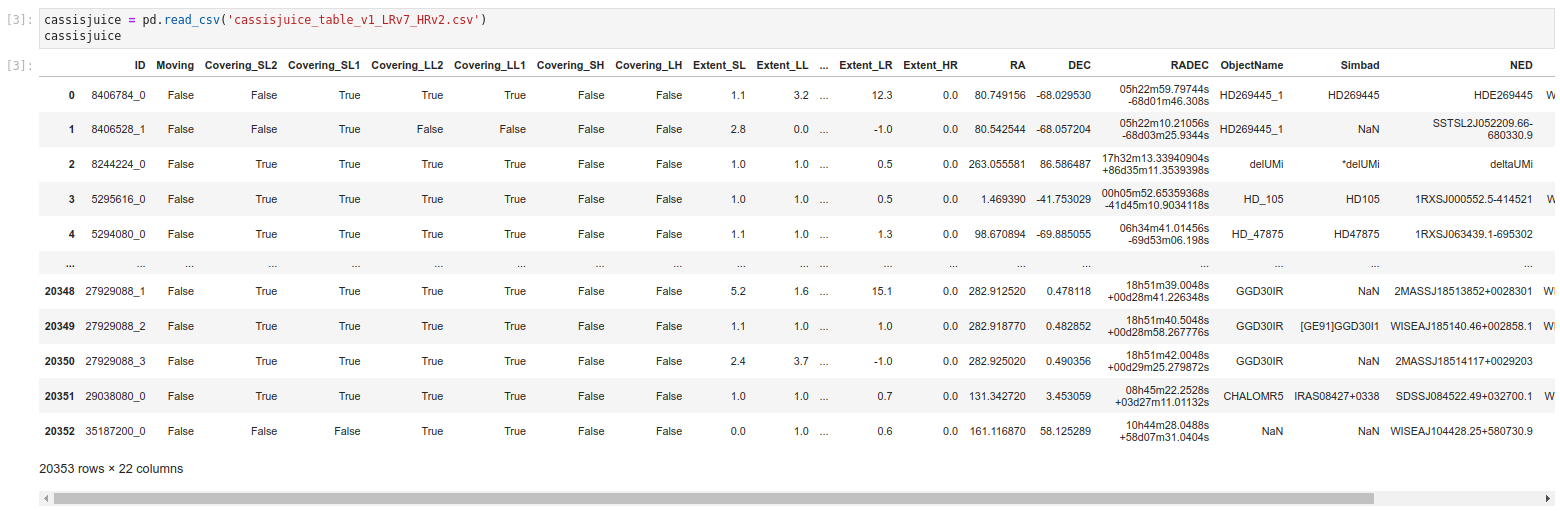}
      \caption{CASSISjuice catalog.}
         \label{fig:table}
       \end{figure*}

From the pipeline described above, we produced a single table (Fig.\,\ref{fig:table}) with each entry defined as a spectral dataset corresponding to a given targeted position either in ``single'' staring observation (i.e., one position per AORkey) or within a ``cluster'' staring observation (i.e., several positions/pointings per AORkey). The CASSISjuice ID is thus fully defined by the AORKey + a number corresponding to the pointing. For each ID, the table provides the extracted coordinates as well as resolved SIMBAD and NED object names, and potentially NED redshift if relevant. Each ID therefore corresponds to a single spectral dataset with either LR or HR spectra or both.

\begin{figure}
   \centering
   \includegraphics[width=9cm]{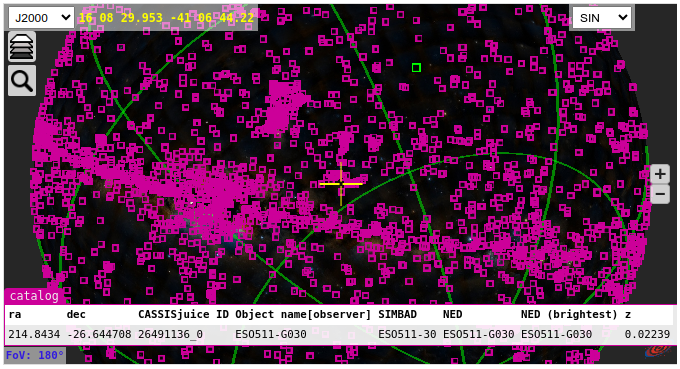}
   \includegraphics[width=9cm]{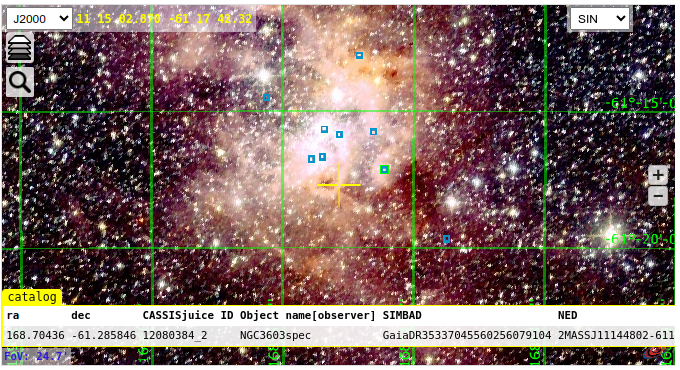}
      \caption{All-sky map with the CASSISjuice catalog (top) and zoom on a specific region (here NGC\,3603; bottom) using Aladin.}
         \label{fig:aladin}
       \end{figure}

Figure\,\ref{fig:aladin} shows the all-sky catalog of CASSISjuice data using Aladin \citep{Baumann2022a} as presented in the Python notebooks within the GitLab repository. The atlas repository also includes convenience scripts to find observations using an object name or coordinates as constraints. (see applications in Sect.\,\ref{sec:apps}) 

\subsection{Spectral dataset and ID cards}

\begin{figure*}
   \centering
   \includegraphics[width=18cm]{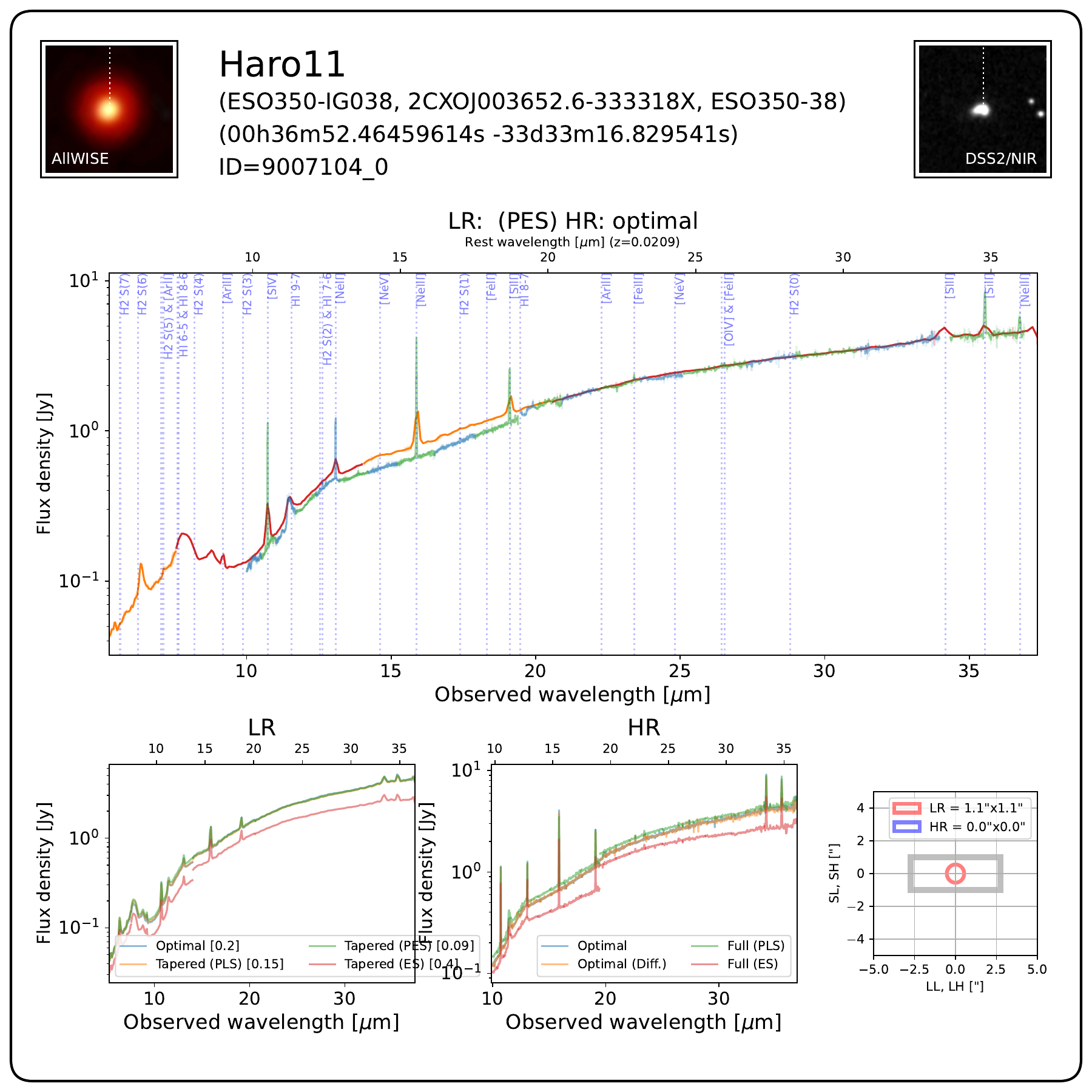}
      \caption{Example of an ID card for a given CASSISjuice ID (here 9007104\_0). The header shows sky snapshots with the targeted position, the object name given by the observer, the name resolved by SIMBAD and NED, the coordinates and the catalog ID. The main spectrum shows the best extraction chosen by the pipeline (here tapered column extraction with partially-extended source calibration for LR and optimal extraction for HR), with this choice partly controlled by the spatial extent in each module (bottom right). The bottom spectra show the various versions for LR and HR spectra. } 
         \label{fig:idcard}
       \end{figure*}

Each CASSISjuice ID is associated with a spectral dataset formatted as a binary product with metadata. Notebooks are provided to read and manipulate these files and potentially export them in other formats. In addition, we also propose a synthetic ID card that encompasses most of the useful diagnostics to compare the various methods for spectral extraction described in Section\,\ref{sec:pipeline}. An example of such ID cards is shown in Figure\,\ref{fig:idcard}.

\subsection{Archives}

We provide two versions of the CASSISjuice atlas for downloading, with the goal in mind to ensure the durability of the IRS spectral atlas and to make it possible for users to work easily on the full database, e.g., through blind scans. 

\begin{figure*}
   \centering
   \includegraphics[width=18cm]{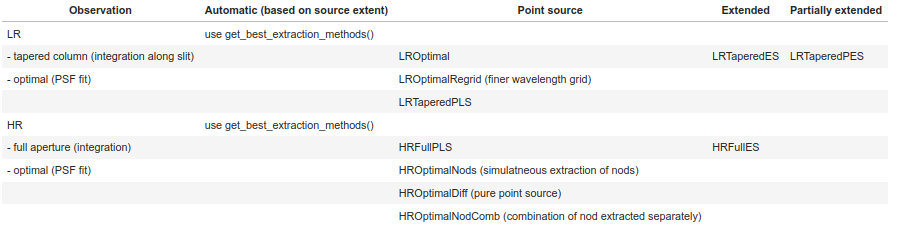}
      \caption{Keys corresponding to the various versions of spectra.}
         \label{fig:keys}
       \end{figure*}

The first atlas version contains the various spectra that use different background subtraction, extraction, and flux calibration methods. The labels for the various methods are provided in Fig.\,\ref{fig:keys} and the various resulting spectra can be examined manually or, for instance, with the help of the ID card. 

\begin{figure}
   \centering
   \includegraphics[width=9cm]{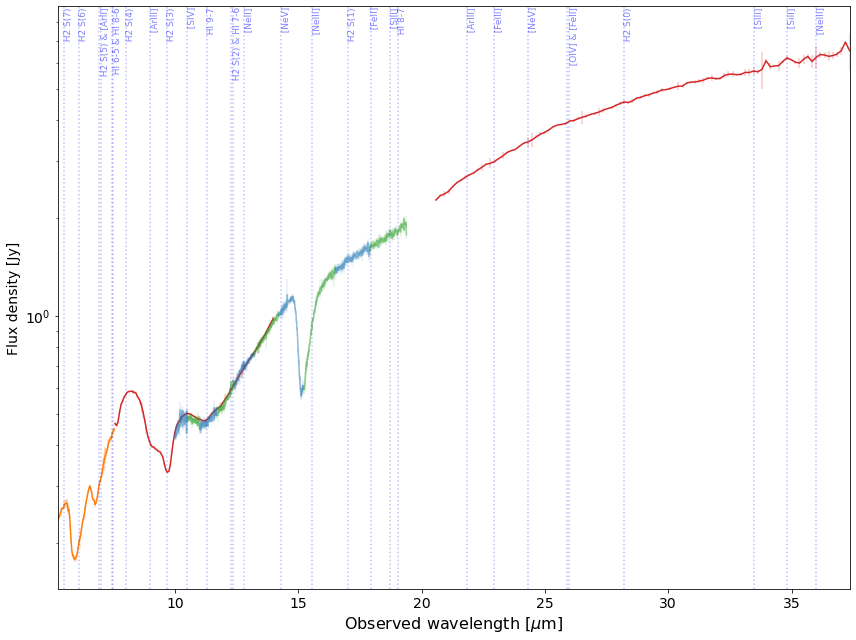}
      \caption{Example of the full LR+HR spectral dataset plotted for one CASSISjuice ID (here 14897920\_0 for which the LL spectral order 2 is missing but is partly covered by SH).} 
         \label{fig:specfull}
       \end{figure}

The second atlas version (``concentrate'') contains only the default spectrum for each ID, i.e., following the pipeline automatic decision tree for the various methods mentioned above. The choice of the spectral extraction method depends on the source spatial extent, leading to a single spectral dataset for a given ID (see example in Fig.\,\ref{fig:specfull}). CASSISjuice concentrate is meant to provide a simple and small archive of all IRS staring observations for users who do not need to confirm/change the automatic selection for the best spectra. However, we do encourage a systematic comparison of the various methods when possible.

\section{Applications and illustrations}\label{sec:apps}

The online repositories provide several Python notebooks to illustrate how the CASSISjuice atlas may be used and manipulated. Apart from the obvious applications to obtain spectra of specific objects (as illustrated in Fig.\,\ref{fig:idcard}), having access to the entire spectral database makes it possible to perform global calculations in order to, e.g., find spectra reassembling others, find spectra with a particular feature, combine spectra to produce templates etc...

\subsection{Finding spectra}\label{sec:finding}

The basic way to find spectra relies either on a set of coordinates or on object names. The notebooks in the online repositories show example in each case. For object names, the match can be performed on the object name given by the observer and/or on the object name resolved from SIMBAD or NED at the observed coordinates.

\begin{figure}
   \centering
   \includegraphics[width=9cm]{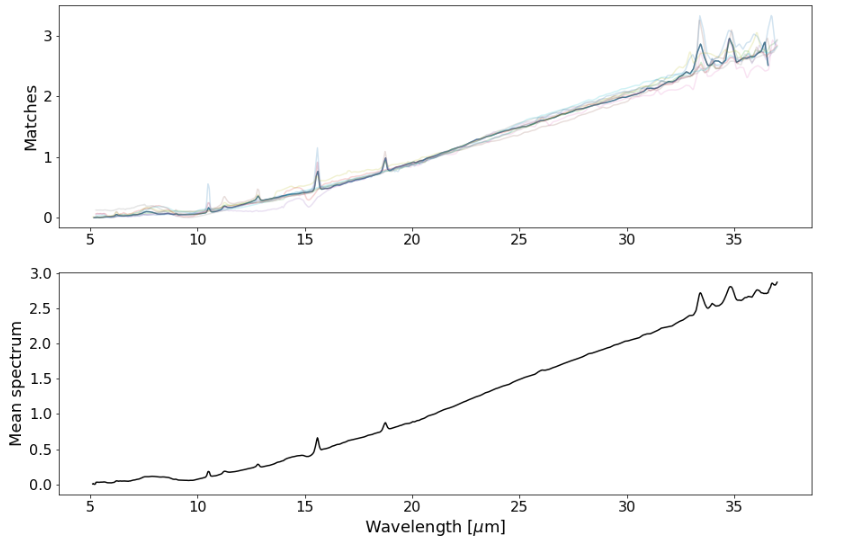}
      \caption{Illustration of a blind scan to identify spectra resembling best a given reference spectrum (here considering the full wavelength range). The top panel shows the reference spectrum in black and the first best matches in color, while the bottom panel shows the mean spectrum of the best matches.} 
         \label{fig:similar}
       \end{figure}

\begin{figure}
   \centering
   \includegraphics[width=9cm]{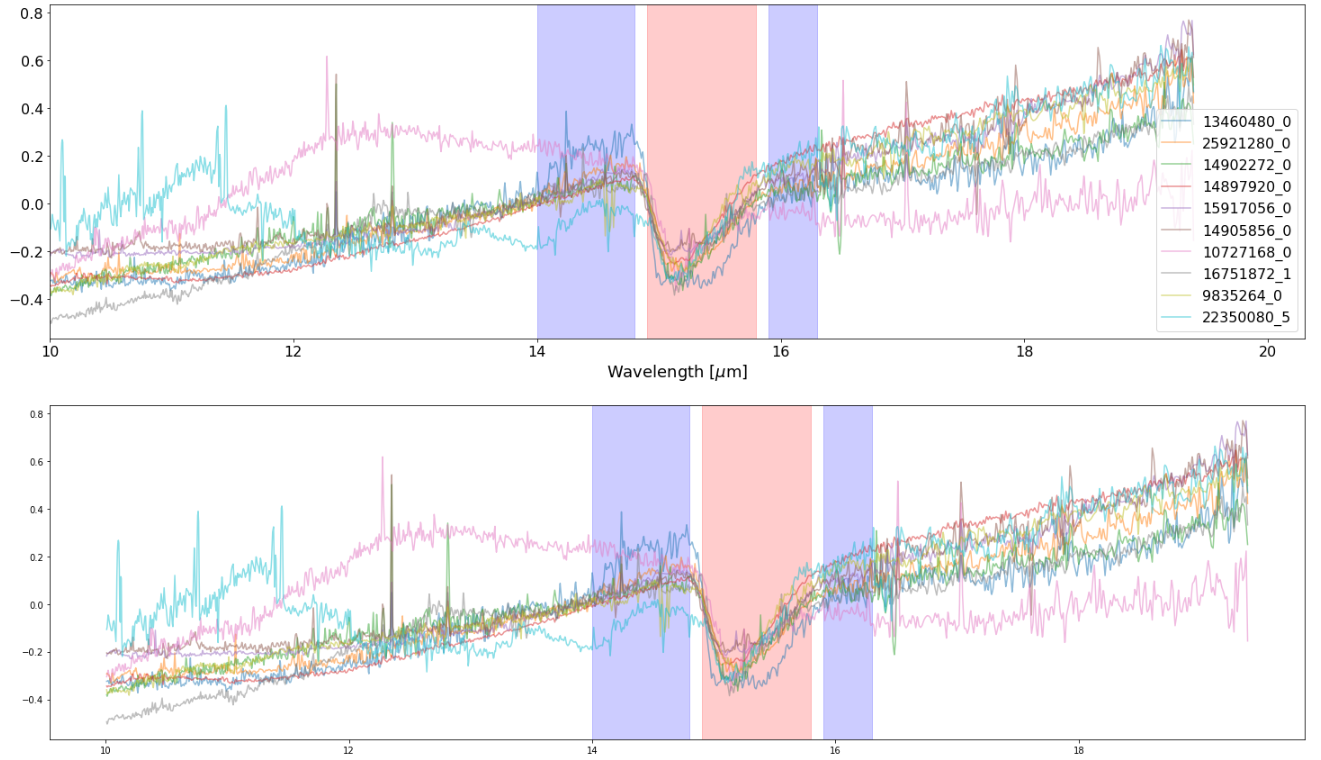}
      \caption{Illustration of a blind scan to identify spectra based on some constraints on a particular feature measured on-the-fly (here looking for absorption around $15.2$\mic).} 
         \label{fig:otfflux}
       \end{figure}

Another way to find spectra relies on the spectra themselves and some particular features. For instance, from a reference spectrum (be it a model, a JWST spectrum, or even another CASSISjuice spectrum), it is often useful to identify all spectra that show a resemblance, either for the full spectrum or for a given wavelength range (see example in Fig.\,\ref{fig:similar}). Another method consists in measuring some features on-the-fly and select the spectra that match some constraints (see example in Fig.\,\ref{fig:otfflux}). Such blind scans may reveal observations that have been ignored, e.g., due to cryptic object names or to the lack of systematic search for a specific feature.

\subsection{Templates}

\begin{figure*}
   \centering
   \includegraphics[width=18cm]{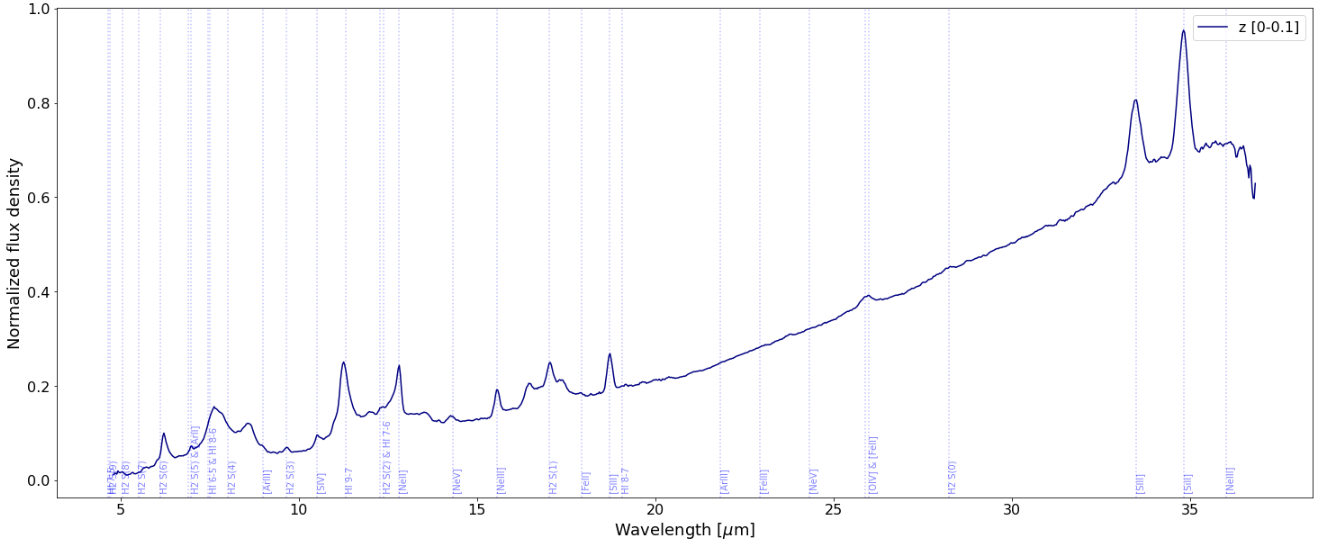}
      \caption{Low-resolution template for extragalactic spectra at redshift $\approx0$.}
         \label{fig:LR_template}
       \end{figure*}

\begin{figure*}
   \centering
   \includegraphics[width=9cm]{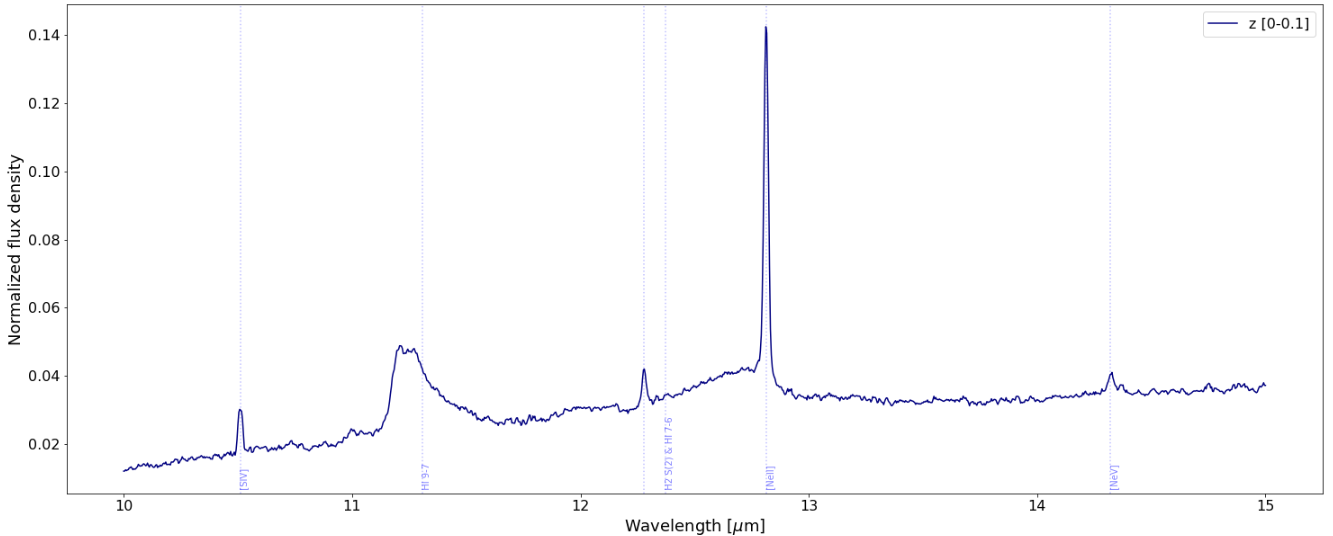}
   \includegraphics[width=9cm]{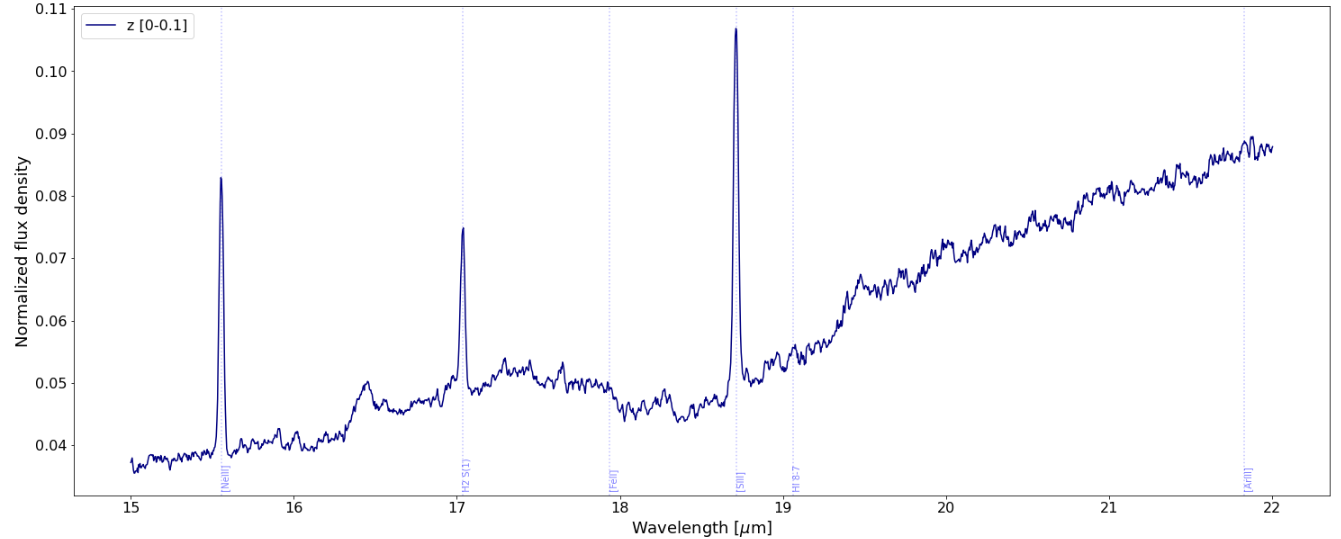}
   \includegraphics[width=9cm]{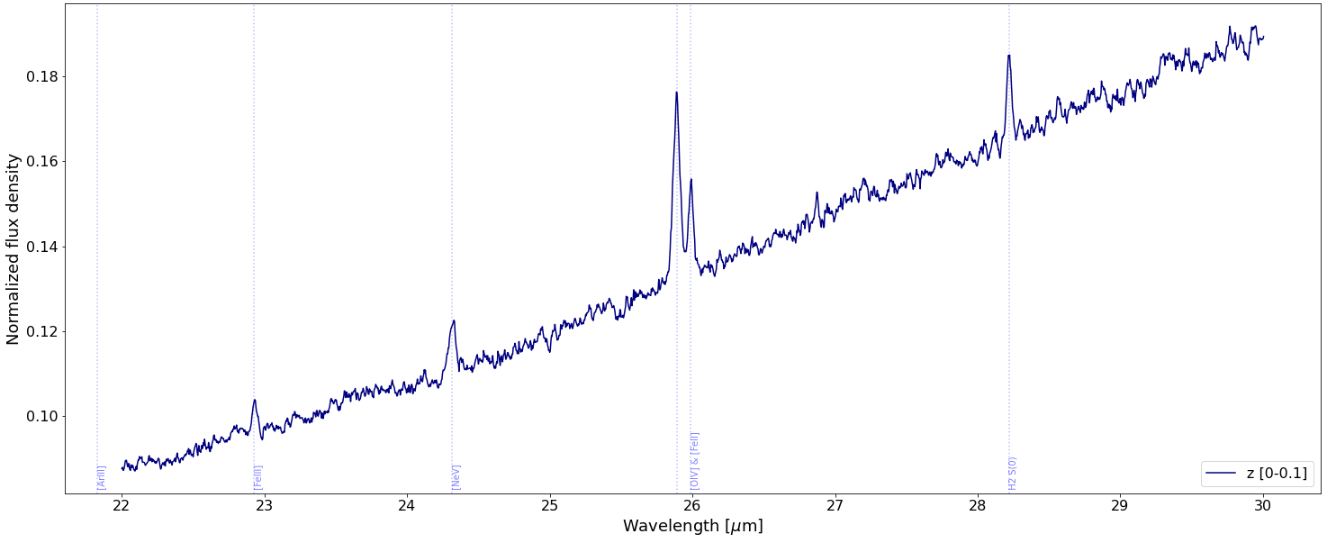}
   \includegraphics[width=9cm]{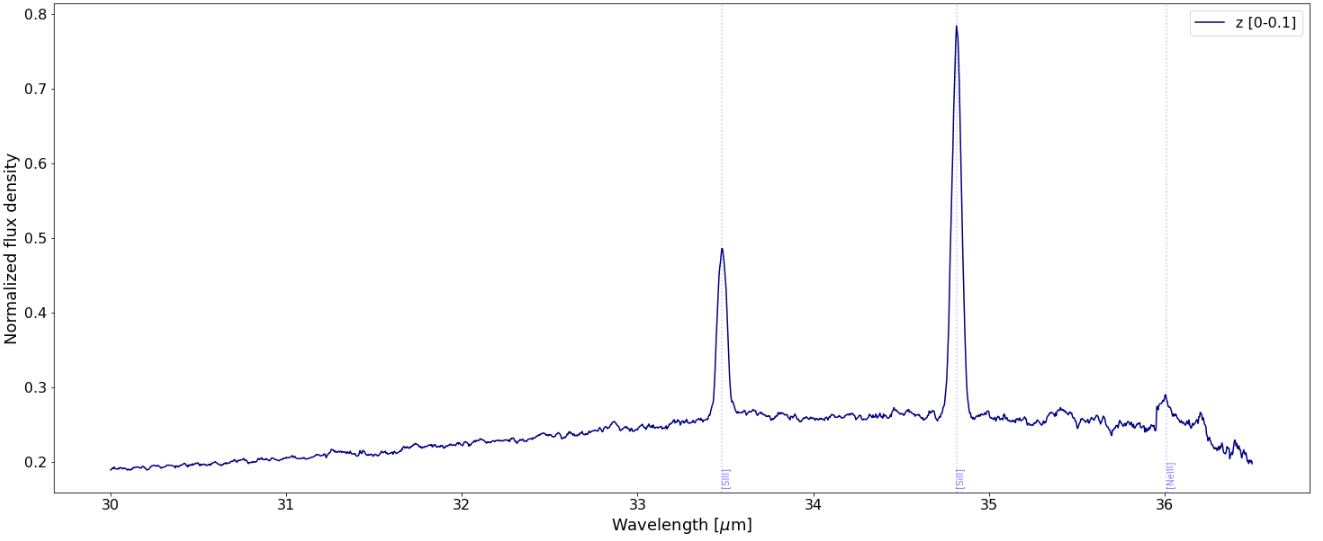}
      \caption{High-resolution template for extragalactic spectra at redshift $\approx0$.}
         \label{fig:HR_template}
       \end{figure*}

From a pre-defined sample or from a sample built from observational constraints/parameters (for instance through spectral resemblance or specific features; Sect.\,\ref{sec:finding}), it is straightforward to produce combinations of spectra that may serve as high-SNR templates. We illustrate this application in Figures\,\ref{fig:LR_template} and\,\ref{fig:HR_template} where we show the low- and high-resolution spectral template for extragalactic CASSISjuice IDs (using the redshift from the NED-resolved object) at redshift $\approx0$, with no particular constraints on the object type or spectral shape. 

\subsection{Extragalactic sample}

While the CASSISjuice catalog includes NED-resolved object names and redshifts if applicable, there may be confusion between the various sources within a given radius. The IDEOS project (Infrared Database of Extragalactic Observables with Spitzer; \citealt{HernanCaballero2016a,Spoon2022a}; \url{http://ideos.astro.cornell.edu/}) precisely aimed at compiling a clean set of extragalactic low-resolution spectra and on-the-fly measurements (redshift, spectral feature fluxes...), using the same data as in CASSISjuice. IDEOS provides a database of extragalactic objects with the best spectrum possible (in particular when several observations are available for a given object) and with a specific attention to homogeneity in the spectral feature measurements between the two low-resolution modules. We provide some examples of practical applications of CASSISjuice and IDEOS within the repositories and we refer to \cite{Spoon2022a} for some astrophysical applications.

\section{Going further}

CASSISjuice extracts spectra at the requested position and either integrates the flux for extended sources or fits the PSF profile for point sources. In some cases the extended object shows a spatial structure indicative of several components with a potentially different spectral shape for each component. In such cases, we advise the use of the SMART/AdOpt software \citep{Lebouteiller2010a} which is specifically built to simultaneously extract spatial components in low-resolution observations. This may correspond, for instance, to supernovae embedded in a galactic disk, to an active galactic nucleus and the host galaxy etc...

%

\section{Conclusions}

\begin{enumerate}
  \item We present CASSISjuice, an open-source, offline version of the pipeline and atlas of Spitzer/IRS observations performed in staring mode (i.e., not including maps).
  \item Repositories are provided for the pipeline code and the atlas itself with the cleanest structure and smallest size possible.
  \item The pipeline has been updated in order to process all observations performed with the IRS, adding a few hundred sources. The high-resolution pipeline has been upgraded to version 2.
    \item One CASSISjuice ID corresponds to one targetted position, with a specific attention to associate the low- and high-resolution spectra within ``cluster'' observations.
      \item Two versions of the atlas are provided, one with the various methods for background subtraction, spectral extraction, and flux calibration, and another ``concentrate'' one with only the spectrum corresponding to the best methods chosen automatically by the pipeline. 
      \item CASSISjuice is meant to facilitate the analysis of large samples or to the identification of interesting/relevant spectra taken with the IRS. Several Python notebooks are provided to illustrate the manipulation of the data and of the full atlas in an offline way.
        \item If CASSISjuice data is used, we encourage citing the pipeline version number (currently v7 for LR and v2 for HR) together with the two reference papers \citep{Lebouteiller2011a,Lebouteiller2015a} and possibly the present arXiv-only paper as well. The pipeline code \citep{Lebouteiller2023b} and atlas \citep{Lebouteiller2023c}  themselves may be cited as well as an alternative to simply mentioning the version number.
        \end{enumerate}

        \begin{acknowledgements}
          This work was supported by the Programme National « Physique et Chimie du Milieu Interstellaire » (PCMI) and the « Programme National de Physique Stellaire » (PNPS) of the CNRS/INSU with INC/INP co-funded by CEA and CNES. This research has made use of the NASA/IPAC Infrared Science Archive, which is funded by the National Aeronautics and Space Administration and operated by the California Institute of Technology. This research has made use of "Aladin sky atlas" developed at CDS, Strasbourg Observatory, France. This research has made use of the NASA/IPAC Extragalactic Database, which is funded by the National Aeronautics and Space Administration and operated by the California Institute of Technology. This research has made use of the SIMBAD database, operated at CDS, Strasbourg, France. 
\end{acknowledgements}

%
%

\bibliographystyle{aa} 
\bibliography{/local/home/vleboute/ownCloud/bibtexendum/bibtexendum.bib} 

\end{document}